\PassOptionsToPackage{unicode}{hyperref}
\PassOptionsToPackage{hyphens}{url}
\PassOptionsToPackage{dvipsnames,svgnames,x11names}{xcolor}
\documentclass[10pt,a4paper,onecolumn]{article}
\usepackage{marginnote}
\usepackage[rgb,svgnames]{xcolor}
\usepackage{authblk,etoolbox}
\usepackage{titlesec}
\usepackage{calc}
\usepackage{tikz}
\usepackage[pdfa]{hyperref}
\usepackage{hyperxmp}
\hypersetup{%
    unicode=true,
    pdfapart=3,
    pdfaconformance=B,
    pdftitle={Shear Mapping in Python (SMPy): Modular, Extensible, and
Accessible Dark Matter Mapping},
    pdfauthor={Georgios N. Vassilakis, Jacqueline E. McCleary, Maya
Amit, Sayan Saha},
    pdfpublication={Journal of Open Source Software},
    pdfpublisher={Open Journals},
    pdfissn={2475-9066},
    pdfpubtype={journal},
    pdfvolumenum={},
    pdfissuenum={},
    pdfdoi={10.21105/joss.10928},
    pdfcopyright={Copyright (c) 2026, Georgios N. Vassilakis, Jacqueline
E. McCleary, Maya Amit, Sayan Saha},
    pdflicenseurl={http://creativecommons.org/licenses/by/4.0/},
    colorlinks=true,
    linkcolor=[rgb]{0.0, 0.5, 1.0},
    citecolor=Blue,
    urlcolor=[rgb]{0.0, 0.5, 1.0},
    breaklinks=true
}
\immediate\pdfobj stream attr{/N 3} file{sRGB.icc}
\pdfcatalog{%
  /OutputIntents [
    <<
      /Type /OutputIntent
      /S /GTS_PDFA1
      /DestOutputProfile \the\pdflastobj\space 0 R
      /OutputConditionIdentifier (sRGB)
      /Info (sRGB)
    >>
  ]
}
\usepackage{caption}
\usepackage{orcidlink}
\usepackage{tcolorbox}
\usepackage{amssymb,amsmath}
\usepackage{ifxetex,ifluatex}
\usepackage{seqsplit}
\usepackage{xstring}

\makeatletter
\@ifundefined{KOMAClassName}{
  \IfFileExists{parskip.sty}{%
    \usepackage{parskip}
  }{
    \setlength{\parindent}{0pt}
    \setlength{\parskip}{6pt plus 2pt minus 1pt}}
}{
  \KOMAoptions{parskip=half}}
\makeatother
\usepackage{graphicx}
\makeatletter
\newsavebox\pandoc@box
\newcommand*\pandocbounded[1]{
  \sbox\pandoc@box{#1}%
  \Gscale@div\@tempa{\textheight}{\dimexpr\ht\pandoc@box+\dp\pandoc@box\relax}%
  \Gscale@div\@tempb{\linewidth}{\wd\pandoc@box}%
  \ifdim\@tempb\p@<\@tempa\p@\let\@tempa\@tempb\fi
  \ifdim\@tempa\p@<\p@\scalebox{\@tempa}{\usebox\pandoc@box}%
  \else\usebox{\pandoc@box}%
  \fi%
}
\def\fps@figure{htbp}
\makeatother
\NewDocumentCommand\citeproctext{}{}
\NewDocumentCommand\citeproc{mm}{%
  \begingroup\def\citeproctext{#2}\cite{#1}\endgroup}
\makeatletter
 \let\@cite@ofmt\@firstofone
 \def\@biblabel#1{}
 \def\@cite#1#2{{#1\if@tempswa , #2\fi}}
\makeatother
\newlength{\cslhangindent}
\newlength{\csllabelwidth}
\newenvironment{CSLReferences}[2] 
 {\begin{list}{}{%
  \setlength{\itemindent}{0pt}
  \setlength{\leftmargin}{0pt}
  \setlength{\parsep}{0pt}
  \ifodd #1
   \setlength{\leftmargin}{\cslhangindent}
   \setlength{\itemindent}{-1\cslhangindent}
  \fi
  \setlength{\itemsep}{#2\baselineskip}}}
 {\end{list}}
\usepackage{calc}

\ifLuaTeX
\usepackage[bidi=basic,shorthands=off,]{babel}
\else
\usepackage[bidi=default,shorthands=off,]{babel}
\fi
\ifLuaTeX
  \usepackage{selnolig} 
\fi
\providecommand{\tightlist}{%
  \setlength{\itemsep}{0pt}\setlength{\parskip}{0pt}}

\usepackage{float}
\let\origfigure\figure
\let\endorigfigure\endfigure
\renewenvironment{figure}[1][2] {
    \expandafter\origfigure\expandafter[H]
} {
    \endorigfigure
}

\usepackage[top=3.5cm, bottom=3cm, right=1.5cm, left=1.0cm,
            headheight=2.2cm, reversemp, includemp, marginparwidth=4.5cm]{geometry}

\titleformat{\section}
  {\normalfont\sffamily\Large\bfseries}
  {}{0pt}{}
\titleformat{\subsection}
  {\normalfont\sffamily\large\bfseries}
  {}{0pt}{}
\titleformat{\subsubsection}
  {\normalfont\sffamily\bfseries}
  {}{0pt}{}
\titleformat*{\paragraph}
  {\sffamily\normalsize}

\usepackage{fancyhdr}
\makeatletter
\let\ps@plain\ps@fancy
\fancyheadoffset[L]{4.5cm}
\fancyfootoffset[L]{4.5cm}

\definecolor{linky}{rgb}{0.0, 0.5, 1.0}

\newtcolorbox{repobox}
   {colback=red, colframe=red!75!black,
     boxrule=0.5pt, arc=2pt, left=6pt, right=6pt, top=3pt, bottom=3pt}

\newcommand{\ExternalLink}{%
   \tikz[x=1.2ex, y=1.2ex, baseline=-0.05ex]{%
       \begin{scope}[x=1ex, y=1ex]
           \clip (-0.1,-0.1)
               --++ (-0, 1.2)
               --++ (0.6, 0)
               --++ (0, -0.6)
               --++ (0.6, 0)
               --++ (0, -1);
           \path[draw,
               line width = 0.5,
               rounded corners=0.5]
               (0,0) rectangle (1,1);
       \end{scope}
       \path[draw, line width = 0.5] (0.5, 0.5)
           -- (1, 1);
       \path[draw, line width = 0.5] (0.6, 1)
           -- (1, 1) -- (1, 0.6);
       }
   }

\definecolor{c53baa1}{RGB}{83,186,161}
\definecolor{c202826}{RGB}{32,40,38}

\patchcmd{\@maketitle}{center}{flushleft}{}{}
\patchcmd{\@maketitle}{center}{flushleft}{}{}
\patchcmd{\@maketitle}{\LARGE}{\LARGE\sffamily}{}{}
\def\maketitle{{%
  
  \AB@maketitle}}
\renewcommand\AB@affilsepx{ \protect\Affilfont}
\renewcommand\AB@affilnote[1]{{\bfseries #1}\hspace{3pt}}
\renewcommand{\affil}[2][]%
   {\newaffiltrue\let\AB@blk@and\AB@pand
      \if\relax#1\relax\def\AB@note{\AB@thenote}\else\def\AB@note{#1}%
        \setcounter{Maxaffil}{0}\fi
        \begingroup
        \let\href=\href@Orig
        \let\protect\@unexpandable@protect
        \def\thanks{\protect\thanks}\def\footnote{\protect\footnote}%
        \@temptokena=\expandafter{\AB@authors}%
        {\def\\{\protect\\\protect\Affilfont}\xdef\AB@temp{#2}}%
         \xdef\AB@authors{\the\@temptokena\AB@las\AB@au@str
         \protect\\[\affilsep]\protect\Affilfont\AB@temp}%
         \gdef\AB@las{}\gdef\AB@au@str{}%
        {\def\\{, \ignorespaces}\xdef\AB@temp{#2}}%
        \@temptokena=\expandafter{\AB@affillist}%
        \xdef\AB@affillist{\the\@temptokena \AB@affilsep
          \AB@affilnote{\AB@note}\protect\Affilfont\AB@temp}%
      \endgroup
       \let\AB@affilsep\AB@affilsepx
}
\makeatother

\renewcommand\Affilfont{\sffamily\small\mdseries}
\usepackage[T1]{fontenc}
\usepackage[utf8]{inputenc}
\usepackage{textcomp}
\usepackage{lmodern}

\ifPDFTeX\else
\fi
\IfFileExists{upquote.sty}{\usepackage{upquote}}{}
\IfFileExists{microtype.sty}{
  \usepackage[]{microtype}
  \UseMicrotypeSet[protrusion]{basicmath} 
}{}

\usepackage[varqu]{zi4}

\PassOptionsToPackage{usenames,dvipsnames}{color} 
\ifLuaTeX
  \usepackage{selnolig}  
\fi

\title{Shear Mapping in Python (SMPy): Modular, Extensible, and
Accessible Dark Matter Mapping}

\author[1,2,3,4%
\ensuremath\mathparagraph]{Georgios N. Vassilakis%
  \,\orcidlink{0009-0006-2684-2961}\,%
}
\author[1%
]{Jacqueline E. McCleary%
  \,\orcidlink{0000-0002-9883-7460}\,%
}
\author[1,5%
]{Maya Amit%
  \,\orcidlink{0009-0002-2898-7022}\,%
}
\author[1%
]{Sayan Saha%
  \,\orcidlink{0000-0002-6044-2164}\,%
}

\affil[1]{Department of Physics, Northeastern University, Boston, MA,
USA%
}
\affil[2]{Institute of Astronomy, University of Cambridge, Cambridge,
UK%
}
\affil[3]{Department of Applied Mathematics and Theoretical Physics,
University of Cambridge, Cambridge, UK%
}
\affil[4]{Kavli Institute for Cosmology, University of Cambridge,
Cambridge, UK%
}
\affil[5]{Center for Data Science, New York University, New York, NY,
USA%
}
\affil[$\mathparagraph$]{Corresponding author}
\date{\vspace{-2.5ex}}

\begin{document}
\maketitle

\marginpar{

  \begin{flushleft}
  \sffamily\small

  {\bfseries DOI:} \href{https://doi.org/10.21105/joss.10928}{\color{linky}{10.21105/joss.10928}}

  \vspace{2mm}
    {\bfseries Software}
  \begin{itemize}
    \setlength\itemsep{0em}
    \item \href{https://github.com/openjournals/joss-reviews/issues/10928}{\color{linky}{Review}} \ExternalLink
    \item \href{https://github.com/GeorgeVassilakis/SMPy}{\color{linky}{Repository}} \ExternalLink
    \item \href{https://doi.org/10.5281/zenodo.23710092}{\color{linky}{Archive}} \ExternalLink
  \end{itemize}

  \vspace{2mm}
  
    \par\noindent\hrulefill\par

  \vspace{2mm}

  {\bfseries Editor:} \href{https://github.com/warrickball}{Warrick
Ball} \ExternalLink
   \\
  \vspace{1mm}
    {\bfseries Reviewers:}
  \begin{itemize}
  \setlength\itemsep{0em}
    \item \href{https://github.com/ismael-mendoza}{@ismael-mendoza}
    \item \href{https://github.com/ahuang314}{@ahuang314}
    \end{itemize}
    \vspace{2mm}
  
    {\bfseries Submitted:} 14 May 2025\\
    {\bfseries Published:} 28 October 2026

  \vspace{2mm}
  {\bfseries License}\\
  Authors of papers retain copyright and release the work under a Creative Commons Attribution 4.0 International License (\href{https://creativecommons.org/licenses/by/4.0/}{\color{linky}{CC BY 4.0}}).

  \end{flushleft}
}

\section{Summary}\label{summary}

Understanding the Universe's large-scale distribution of dark matter is
a central objective in the era of precision cosmology. A key technique
for the study of dark matter is weak gravitational lensing: a phenomenon
where light from distant galaxies is sheared as it passes through the
gravitational field of a massive object, like a galaxy cluster. This
shear, which manifests as a slight (weak) distortion of shapes over
thousands of galaxies, allows astrophysicists to infer the distribution
of total matter, including both luminous and dark matter.

Obtaining a mass distribution from a catalog of galaxy shears involves
reconstructing the convergence (\(\kappa\)), which describes how
gravitational lensing changes the apparent sizes of background galaxies.
Convergence is the projected surface mass density divided by the
critical surface density for the lens-source system, allowing it to
trace the projected mass distribution. For a comprehensive review of
weak gravitational lensing, refer to Umetsu
(\citeproc{ref-Umetsu2020}{2020}).

The \textbf{Shear Mapping in Python (\texttt{SMPy})} package provides a
standardized, well-documented, and open-source solution for creating
convergence maps from weak lensing galaxy shear measurements.
\texttt{SMPy} was initially developed to support the Superpressure
Balloon-borne Imaging Telescope (SuperBIT), a stratospheric, near-UV to
near-IR observing platform which completed its 45-night observing run in
spring 2023 with over 30 galaxy cluster observations
(\citeproc{ref-Gill2024}{Gill et al., 2024};
\citeproc{ref-Sirks2023}{Sirks et al., 2023}). \texttt{SMPy} has since
evolved into a general-purpose tool suitable for analyzing the weak
lensing data from cosmological surveys.

\section{Statement of need}\label{statement-of-need}

While mass maps are a key deliverable of many cosmological analyses
(\citeproc{ref-DESY32021}{Jeffrey et al., 2021};
\citeproc{ref-ACTDR62024}{Madhavacheril et al., 2024};
\citeproc{ref-HSC2017}{Oguri et al., 2018}), scientists are often left
to make these maps from scratch. \texttt{SMPy} addresses an outstanding
need for the lensing community: an accessible, well-documented, and
extensible tool to construct publication-quality mass maps from galaxy
shear data. Built on standard scientific Python packages, it provides an
easy entry point for researchers new to mass mapping, while also being
robust for more senior scientific use.

There are currently three separate mapping methods implemented into
\texttt{SMPy}: the classic Kaiser--Squires inversion
(\citeproc{ref-KS1993}{Kaiser \& Squires, 1993}), aperture mass mapping
(\citeproc{ref-Leonard2012}{Leonard et al., 2012};
\citeproc{ref-McCleary2020}{McCleary et al., 2020}), and notably, to our
knowledge, the first publicly available implementation of the full
\texttt{KS+} algorithm (\citeproc{ref-Pires2020}{Pires et al., 2020}).
\texttt{KS+} improves reconstruction quality by correcting for
systematic effects including missing data, field borders, and reduced
shear. \texttt{SMPy} also offers specialized and unique features
valuable for mass mapping, such as flexible coordinate system support
(both celestial and pixel space) and comprehensive signal-to-noise
analysis with multiple noise randomization techniques. Figure
\ref{fig:mass_maps} shows maps of a simulated SuperBIT galaxy cluster
observation made with each of the three methods. \texttt{SMPy} is, to
our knowledge, the first convergence mapping software to prioritize both
accessibility and advanced features.

\section{State of the field}\label{state-of-the-field}

The weak lensing community is served by publicly available mapping tools
like \texttt{lenspack} (\citeproc{ref-lenspack}{Peel, 2020}) and
\texttt{jax-lensing} (\citeproc{ref-Remy2022}{Remy et al., 2023}), each
with their own strengths. \texttt{jax-lensing} excels at
neural-network-based approaches and deep learning methods, but applying
it to survey-specific data products requires substantial additional
development by the end user. \texttt{lenspack} provides well-documented,
stand-alone mass-mapping functions, but does not provide an end-to-end
convergence-mapping workflow (e.g., configuration management, coordinate
handling, noise/SNR analysis, and plotting). In practice, many
mass-mapping analyses still rely on bespoke codes built for a specific
survey or science case. These gaps, combined with the need for a
flexible, configuration-driven, and modular mapping framework, motivated
the development of \texttt{SMPy} as an accessible and extensible
alternative.

\section{Software design}\label{software-design}

\begin{figure}
\centering
\pandocbounded{\includegraphics[keepaspectratio,alt={E-mode maps of a simulated SuperBIT observation of a galaxy cluster (McCleary et al., 2023) made with the three methods in SMPy: Kaiser--Squires and KS+ convergence, aperture mass, and the Kaiser--Squires signal-to-noise map.}]{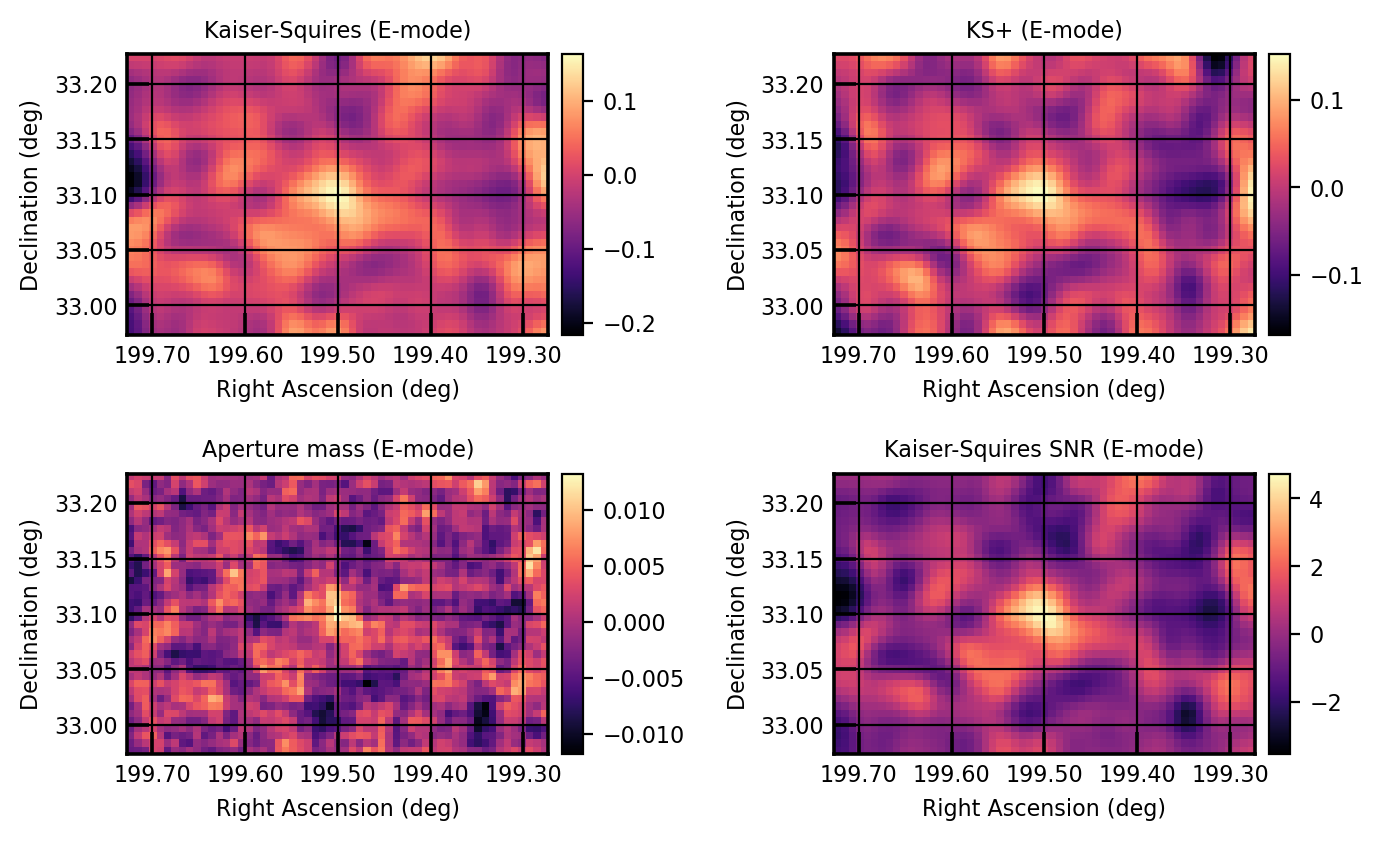}}
\caption{E-mode maps of a simulated SuperBIT observation of a galaxy
cluster (\citeproc{ref-McCleary2023}{McCleary et al., 2023}) made with
the three methods in \texttt{SMPy}: Kaiser--Squires and \texttt{KS+}
convergence, aperture mass, and the Kaiser--Squires signal-to-noise
map.}\label{fig:mass_maps}
\end{figure}

\begin{figure}
\centering
\pandocbounded{\includegraphics[keepaspectratio,alt={Errors of the Kaiser--Squires and KS+ maps made from the noiseless shear of the cluster in Figure , with each map's mean removed. Top: the true convergence and each map's error, smoothed like the maps. Bottom: the same without smoothing, with the galaxies within 1.2 arcmin of eight random points removed (green outlines). KS+ removes the Kaiser--Squires error along the field border.}]{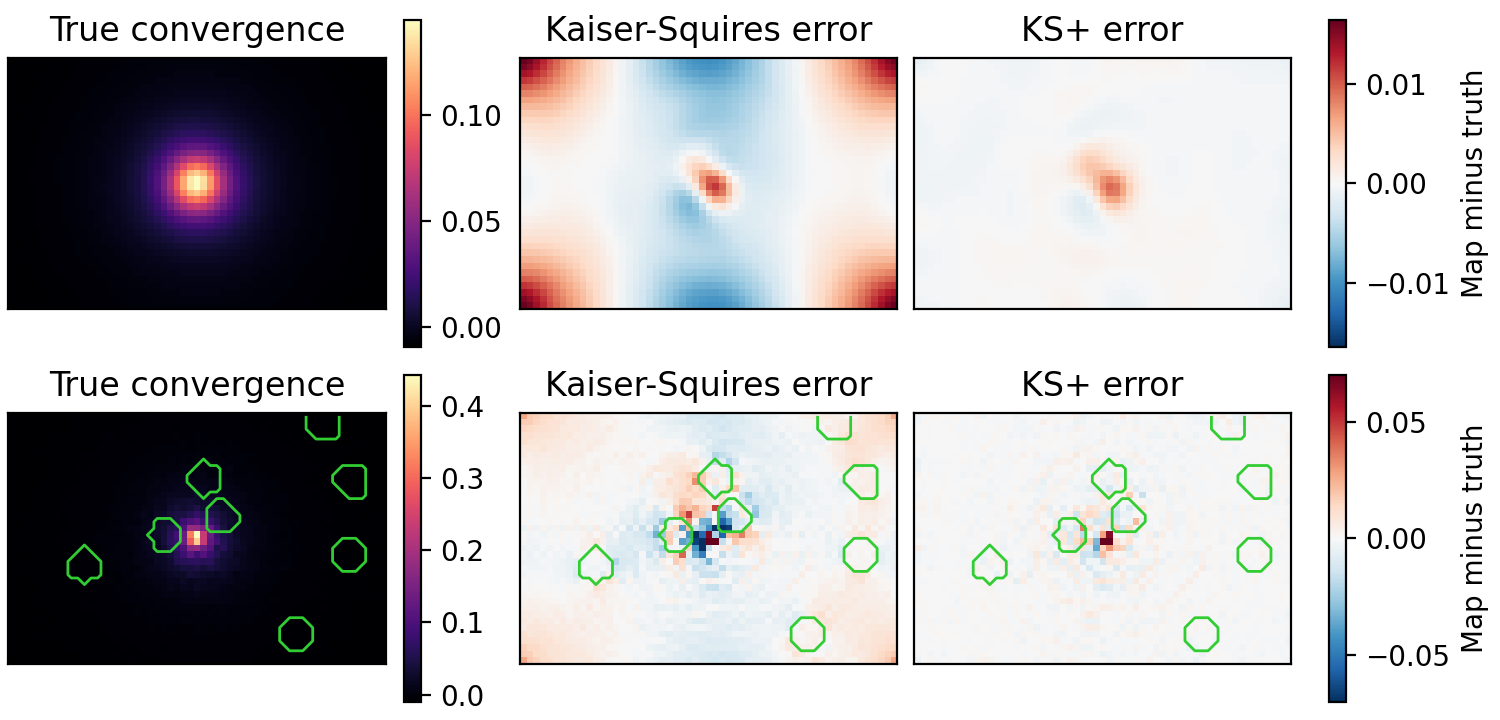}}
\caption{Errors of the Kaiser--Squires and \texttt{KS+} maps made from
the noiseless shear of the cluster in Figure \ref{fig:mass_maps}, with
each map's mean removed. Top: the true convergence and each map's error,
smoothed like the maps. Bottom: the same without smoothing, with the
galaxies within 1.2 arcmin of eight random points removed (green
outlines). \texttt{KS+} removes the Kaiser--Squires error along the
field border.}\label{fig:residuals}
\end{figure}

\texttt{SMPy} was built with the following design principles in mind:

\begin{enumerate}
\def\labelenumi{\arabic{enumi}.}
\item
  \textbf{Accessibility:} \texttt{SMPy} is written entirely in Python
  and deliberately relies only on widely-used scientific Python packages
  (NumPy, SciPy, pandas, Astropy, Matplotlib, and PyYAML). This choice
  of standard dependencies ensures that users can easily install the
  packages without complex dependency chains, and that the codebase is
  maintainable and familiar to the scientific Python community.
\item
  \textbf{Extensibility:} \texttt{SMPy}'s modular architecture enables
  seamless addition of new mass mapping techniques, encouraging
  open-source contribution. The framework's standardized abstract base
  class architecture allows for the integration of different mapping
  methods with minimal effort.
\item
  \textbf{Usability:} Creating convergence maps with \texttt{SMPy}
  requires minimal input---users need to only provide a catalog of
  galaxies with their associated shear components and coordinates. This
  straightforward input requirement makes the tool accessible to
  researchers at all levels. A flexible configuration system is
  integrated via a single YAML file that defines file paths, convergence
  map algorithm settings, plotting parameters, and more. With this
  configuration file, the user can create convergence and SNR maps with
  one line, either via terminal or within code.
\item
  \textbf{Robustness:} \texttt{SMPy} is tested against analytical
  lensing relations and simulated catalogs with known convergence. From
  the noiseless shear of the simulated field in Figure
  \ref{fig:mass_maps}, whose cluster is an NFW halo with
  \(M_{\mathrm{200c}} = 4.1 \times 10^{14}\,h^{-1} M_\odot\) at
  \(z = 0.45\) (\citeproc{ref-McCleary2023}{McCleary et al., 2023}),
  Kaiser--Squires and \texttt{KS+} recover the true convergence with
  correlations of 0.971 and 0.9996, respectively. Within holes cut in
  the same field, \texttt{KS+} reduces the RMS reconstruction error by
  56\% relative to Kaiser--Squires (Figure \ref{fig:residuals}). The
  coordinate system abstraction handles both celestial coordinates
  (using a tangent-plane projection with each shear rotated into the map
  frame) and pixel-based coordinates through a unified interface. To
  quantify the significance of the weak lensing detection, multiple
  noise realizations can be generated using either spatial shuffling
  (randomizing galaxy positions while preserving shear values) or
  orientation shuffling (randomizing shear orientations while preserving
  positions). These noise realizations are mapped with the same method
  and settings as the observations to estimate the noise in each pixel.
  The documentation's validation page describes how the simulated
  catalog was made, how Figure \ref{fig:residuals} and the numbers above
  were produced, and the remaining tests.
\end{enumerate}

These design decisions are reflected in \texttt{SMPy}'s overall
workflow: users provide a shear catalog and a single YAML configuration
file, and \texttt{SMPy} produces convergence maps, signal-to-noise maps,
and publication-quality plots with minimal boilerplate. The mapping
framework is implemented through standardized abstract base classes,
allowing different reconstruction methods to be interchanged while
preserving a consistent interface and coordinate-system handling.

On an Apple M4 Pro, a \(1024 \times 1024\) map from 10.5 million
galaxies takes 0.9 s with Kaiser--Squires, 1.8 s with aperture mass, and
506 s with \texttt{KS+}, using default settings. For a cluster-sized
field like that of Figure \ref{fig:mass_maps}, a \(57 \times 38\) map
from 8,767 galaxies, Kaiser--Squires takes 4 ms, aperture mass 8 ms, and
\texttt{KS+} 0.7 s. The higher cost of \texttt{KS+} reflects its 300
iterations on an extended field. Further timing measurements and scaling
details are provided in the documentation's performance page.

\section{Research impact statement}\label{research-impact-statement}

\texttt{SMPy} was initially developed as part of the weak lensing
analysis pipeline for SuperBIT, and it was used to create the first
convergence maps measured from the stratosphere, presented in ``Lensing
in the Blue IV'' (\citeproc{ref-Saha2026}{Saha, Amit, et al., 2026}).
These maps were made from the weak lensing shape catalogs of the 30
galaxy clusters observed on SuperBIT's 2023 science flight, presented in
``Lensing in the Blue III'' (\citeproc{ref-Saha2026shapes}{Saha,
McCleary, et al., 2026}). In addition, \texttt{SMPy} has been presented
at the Jet Propulsion Laboratory, and we have received interest in
exploring integration into Euclid weak-lensing analysis workflows.

\section{AI usage disclosure}\label{ai-usage-disclosure}

Generative AI was used in this work. First, Claude Sonnet 4 (with
Extended Thinking enabled via the Claude.ai web interface) was used to
assist in implementing the \texttt{KS+} algorithm in \texttt{SMPy}. The
author reviewed all AI-assisted code line-by-line against the Algorithm
1 description in Appendix A of Pires et al.
(\citeproc{ref-Pires2020}{2020}) and validated the implementation by
injecting fiducial star masks into data and confirming that the
reconstruction behaved as expected. Second, OpenAI's ChatGPT
(5.0-Thinking and 5.2-Thinking) models have been used for generating
unit tests for the codebase via the Codex CLI. Lastly, both OpenAI's
ChatGPT models and Anthropic's Claude models have been used in the
GitHub App harness to review pull requests and catch bugs.

During the JOSS peer-review revision, Anthropic's Claude Opus 5.5, via
Claude Desktop, assisted with code review and revision, documentation,
and validation/testing scripts. The author made the design decisions,
reviewed and approved each change, and checked all results. The authors
take full responsibility for this work.

\subsection{Software references}\label{software-references}

\texttt{SMPy} is written in Python 3.10+ and uses the following
packages:

\begin{itemize}
\tightlist
\item
  NumPy (\citeproc{ref-numpy}{Harris et al., 2020})
\item
  SciPy (\citeproc{ref-scipy}{Virtanen et al., 2020})
\item
  pandas (\citeproc{ref-pandas}{Pandas Development Team, 2024})
\item
  Astropy (\citeproc{ref-astropyv1}{Astropy Collaboration et al., 2013},
  \citeproc{ref-astropyv2}{2018}, \citeproc{ref-astropyv3}{2022})
\item
  Matplotlib (\citeproc{ref-matplotlib}{Hunter, 2007})
\item
  PyYAML (\citeproc{ref-pyyaml}{Simonov, 2025})
\end{itemize}

\section{Acknowledgements}\label{acknowledgements}

This material is based upon work supported by a Northeastern University
Undergraduate Research and Fellowships PEAK Summit Award.

\section*{References}\label{references}
\addcontentsline{toc}{section}{References}

\protect\phantomsection\label{refs}

\end{document}